%% file: iopjournal-template.tex
\documentclass[12pt]{article}

\usepackage[utf8]{inputenc}
\usepackage[T1]{fontenc}

\usepackage{gensymb}
\usepackage{units}
\usepackage{amsmath}
\usepackage{amssymb}

\usepackage{tikz}
\usetikzlibrary{shapes.geometric, arrows}

\tikzstyle{startstop} = [rectangle, rounded corners, minimum width=1cm, minimum height=0.5cm,
text centered, draw=black, fill=red!30]
\tikzstyle{io} = [trapezium, trapezium left angle=70, trapezium right angle=110, minimum width=1cm, minimum height=0.5cm, text centered, draw=black, fill=blue!30]
\tikzstyle{process} = [rectangle, minimum width=1cm, minimum height=0.5cm, text centered, draw=black, fill=orange!30]
\tikzstyle{decision} = [diamond, minimum width=0.5cm, minimum height=0.5cm, text centered, draw=black, fill=gray!30]
\tikzstyle{arrow} = [thick,->,>=stealth]

\usepackage{setspace}

\usepackage{url}

\begin{document}

\begin{center}
  {\Large\bfseries Surrogate Models studies for laser-plasma accelerator electron source design}\\[2pt]
  {\Large\bfseries through numerical optimisation}\\[12pt]

  G. Kane,\textsuperscript{a)} P. Drobniak, S. Kazamias, V. Kubytskyi, M. Lenivenko, B. Lucas, J. Serhal, and K. Cassou\\[8pt]

  {\small
  Laboratoire de Physique des 2 Infinis Ir\`ene Joliot-Curie - IJCLab - UMR9012 - B\^at. 100 - 15 rue Georges Cl\'emenceau 91405 Orsay cedex - France.}\\[8pt]

  A. Beck and A. Specka\\[4pt]
  {\small
  Laboratoire Leprince-Ringuet - LLR – UMR 7638 CNRS, \'Ecole Polytechnique, 91128 Palaiseau cedex – France}\\[8pt]

  F. Massimo\\[4pt]
  {\small
  Laboratoire de Physique des Gaz et des Plasmas - LPGP - UMR 8578, CNRS, Universit\'e Paris-Saclay, 91405 Orsay, France}\\[10pt]

  (Dated: 14 October  2025)
\end{center}

\vspace{8pt}


\begin{abstract}
Designing a high-quality plasma injector electron source driven by a laser beam relies on numerical parametric studies using particle-in-cell codes. 
The common input parameters to explore are laser characteristics, plasma species and density profiles produced by computational fluid dynamic studies. We demonstrate the construction of surrogate models using machine learning techniques for a laser-plasma injector (LPI)  based on more than $3000$ particle-in-cell simulations of laser wakefield acceleration performed for sparsely spaced input parameters published by Drobniak [Phys. Rev. Accel. Beams, 26, 091302, (2023)].
Surrogate models are relevant for LPI design and optimisation, as they are approximately $10^7$ times faster than PIC simulations. Their speed enables more efficient design studies by allowing extensive exploration of the input-output relationship without significant computational cost. We develop and compare the performance of three surrogate models, namely, multilayer perceptron (MLP), decision trees (DT) and Gaussian processes (GP). We show that using a simple and frugal MLP-based model trained on a reasonable-size random scan data set of 500 particles in cell simulations, we can predict beam parameters with a coefficient determination score $R^2=0.93$ . The best surrogate model is used to quickly find optimal working points and stability regions and get targeted electron beam energy, charge, energy spread and emittance using different methods, namely random search,  Bayesian optimisation and multi-objective Bayesian optimisation. This simple approach can serve more global design study of an LPI  in a start-to-end linear laser-driven accelerator.
\end{abstract}

\section{Introduction}

Laser wakefield acceleration (LWFA)\cite{tajima1979laser} is a promising method that can produce high-energy electrons 
within compact structures. It can achieve peak accelerating electric field in the order of $100\,$GV/m, $3$ order of magnitude 
higher than the fields generated by RF accelerators \cite{esarey2009physics}. Furthermore, LWFA produces electrons with 
extremely short pulse duration \cite{faure2004laser}, typically around 10's of femtoseconds. This short electron bunch length is 
particularly advantageous for applications like radiotherapy techniques such as FLASH \cite{vozenin2019biological} and 
the creation of coherent X-rays using free electron laser \cite{helml2017ultrashort}. 
In the past decade, several groups were able to generate electron beams with desired properties such as high energy \cite{gonsalves2019petawatt}, high charge \cite{couperus2017demonstration}, low energy spread \cite{ke2021near}, low emittance \cite{mao2015highly}. 
However, these electron beams may not display all these properties simultaneously.
This is due to the highly non-linear and coupled nature of the laser wakefield interaction, making it difficult to obtain a stable electron beam with demanding features.

The nonlinear nature of LWFA makes numerical modelling such as particle-in-cell (PIC) simulations \cite{birdsall2018plasma} necessary for designing reliable laser-plasma accelerators, which can be intractable if one relies only on limited experience data points and scaling laws.
Machine learning (ML) techniques \cite{dopp2023data} are increasingly used in LWFA studies and experiments. Recent papers  \cite{jalas2021bayesian,irshad2023pareto} showed that optimal working points can be obtained by using a Bayesian optimisation approach. 
In this article, we construct and evaluate surrogate models (SM), including multilayer perceptron (MLP), decision trees (DT) and Gaussian processes (GP). These SM are used to predict electron beam properties from input configurations of a laser-plasma injector (LPI) . These models were chosen because they are easy to implement and readily available through numerous Python libraries. Furthermore, these frugal models\footnote{Frugal machine learning focuses on developing highly accurate models while minimising data and resource consumption} demonstrated a high prediction performance in numerous non-linear physics problems \cite{carleo2019machine}. From the models considered, we identify that MLP achieves the best performance with a coefficient of determination $R^2=0.97$. Using the SM, we identified stable operation regions with optimal beam parameters surpassing those found in the previous study \cite{drobniak2023fast}. We demonstrate that SM models are $\approx 10^7$ times faster than PIC simulations, making them significantly more efficient for rapidly exploring various configurations of laser-plasma interactions (LPI) enabling their integration into a comprehensive start-to-end simulation framework for advanced laser-plasma-based electron accelerators. In this paper, SM are applied on SMILEI\cite{derouillat2018smilei} PIC simulations, but could be used with any PIC code. Moreover, they could be integrated with experimental data for real-time operation and optimisation. This study is a necessary step towards providing an efficient approach for designing high-quality electron beams in Laser-wakefield-acceleration applications.
 
The article starts with the numerical experiment section II. detailing the setup and input parameters used for simulations. In section III. we present the data sets and discuss the construction of simple model for predicting injection and filtering data. In section IV. we examine different machine learning approaches for surrogate modelling. In section V. we highlight the performance of these models, discuss some optimisation strategies, and compare SM with conventional methods. Concluding remarks summarise the study's impact on LWFA numerical optimisation. 


\section{Numerical experiment} 

The data set used for the SM training comes from PIC simulations 
aiming to deliver electron beams ranging from $150-250\,$ MeV, $30-50\,$pC of charge, an energy spread lower than $5\%$ and an emittance of less than $2\,$mm.mrad as presented in  \cite{drobniak2023fast}. The LASERIX platform at IJCLab provides the laser driver with a power in the range of $40$ to $80\,$TW. The LPI relies on an ionisation injection scheme \cite{chen2012theory} with a plasma target divided into two regions \cite{golovin2015tunable}. The first region comprises a gas mixture of $He$ doped with $N_2$ whose length is $0.6\,$mm. The inner shell electrons of N$^{5+}$ and N$^{6+}$ can be injected in the plasma wakefield. The second region is composed of pure He, $1.2\,$mm long and dedicated to the acceleration of the injected electrons.
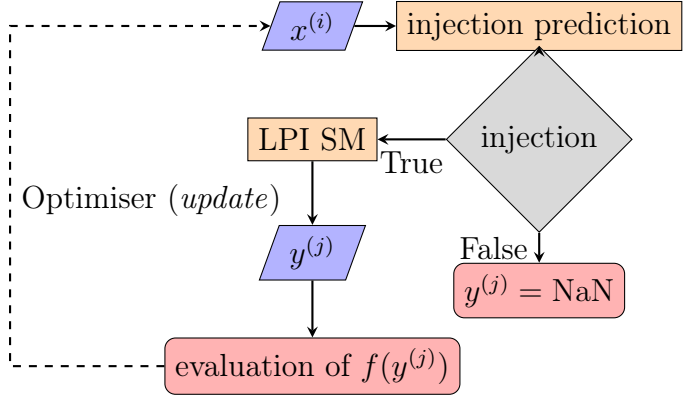
\begin{figure}[!hp]
\centering
\begin{tikzpicture}[node distance=1.5cm]
	\node (in) [io] {$x^{(i)}$}; 
	\node (injSM) [process, right of=in, xshift=1.5cm] {injection prediction};
	\node (injection) [decision, below of=injSM, yshift=-0.0cm] {injection};
	\node (LPISM) [process, left of=injection, xshift= -1.5cm] {LPI SM};
	\node (outoff) [startstop, below of=injection, node distance=2cm] {$y^{(j)} = \mathrm{NaN} $};
	\node (out) [io, below of=LPISM] {$y^{(j)}$}; 
	\node (outopt) [startstop, below of=out] {evaluation of $f(y^{(j)} $)};
	\draw [arrow] (in) -- (injSM);
	\draw [arrow] (injSM) -- (injection);
	\draw [arrow] (injection) -- node[anchor=east] {False} (outoff);
	\draw [arrow] (injection) -- node[anchor=north] {True} (LPISM);
	\draw [arrow] (LPISM) -- (out);
	\draw [arrow] (out) -- (outopt);
	\coordinate[left of=in] (a1);
	\coordinate[left of=outopt] (b1);
	\draw [dashed,arrow] (outopt) -| node [anchor=west,yshift=2.2cm] {Optimiser (\textit{update})} ([xshift=-2.5cm]b1) -- ([xshift=-2.5cm]a1)-- (in);
\end{tikzpicture}
\caption{Surrogate models and injection prediction-based LPI design optimisation studies flowchart. The dashed line represents the potential optimisation loop update of the input $x^{(i)}$.  }
\label{fig:numExp}
\end{figure}
The main objectives of the numerical experiments are: (i) construct a classification-based model that predicts electron beam injection as a function of input parameters; (ii) construct  ML-based SM from simulation data that are able to predict the LPI electron beam parameters; (iii) use both classification model and SM to optimise LPI configurations and investigate the stability of optimal beam parameters.

The scheme in Fig \ref{fig:numExp} illustrates the principle of the numerical experiments. An LPI configuration input $x^{(i)}$ can be filtered by an injection prediction model, providing inputs for the LPI SM to predict the electron beam parameters $y^{(j)}$. An objective function $f$ built from the outputs $y^{(j)}$ can feed an optimisation routine.   


\section{Data sets generation} 

Two large data sets of LPI simulations were produced using the SMILEI\cite{derouillat2018smilei} PIC code with azimuthal mode decomposition, envelope approximation\cite{Massimo2019,Massimo2020cylindrical,massimo2020env} and a low number of macroparticles per cell (MPC). A single run is performed in $130\,$core-hours at the GENCI High-performance computing (HPC) Irene Joliot Curie facility \cite{Genci}, compared to $450\,$core-hours for more standard settings with a higher number of MPC using the envelope and azimuthal mode decomposition. The reduced number of MPC had only a modest impact on the simulation results, as specified in \cite{drobniak2023fast}.
The simulation data are available online \cite{LPISurrogategitlab2024}. The simulations had a set of 4 input variable parameters namely: $x^{(i)} = (a_0,x_{off},p_1,c_{N_2})$ with $a_0$ the laser pulse normalised vector potential in vacuum, $x_{off}$  the laser focal position in vacuum, $p_1$ the gas pressure in the first region and $c_{N_2}$ the concentration of nitrogen in the same region. It is important to notice that the pressure in the second region was kept equal to the one in the first region $p_1 \simeq p_2$, leading to a difference in electron density between the two chambers coming from the 10 electrons of $N$ atoms in the first region. The reference position $x_{off}=0$ corresponds to the entrance of the second chamber \cite{drobniak2023twochambergastargetlaserplasma}. These parameters were selected because they provide a sufficient basis for adjusting and controlling the electron beam parameters in the current design of the LPI project. The input parameters $x^{(i)}$ can vary up to the values indicated in Tab. \ref{tab:inputParameters}. 
 
 \begin{table}[!h]
 \centering
    \begin{tabular}{|c|c|c|c|}
    \hline
    $a_0$ & $x_{off}$ [$\mu$m] & $p_1$ [mbar] & $c_{N_2}$ [$\%$]\\
    \hline
    $[1.1,1.85]$ & $[-400,1800]$ & $[14,119]$ & $[0.2,12]$ \\
    \hline
    \end{tabular}
\caption{Intervals for the four input parameters used for the random scan simulations}
\label{tab:inputParameters}
\end{table}

\begin{figure}[!ht]
    \centering
    \includegraphics[width=0.5\columnwidth]{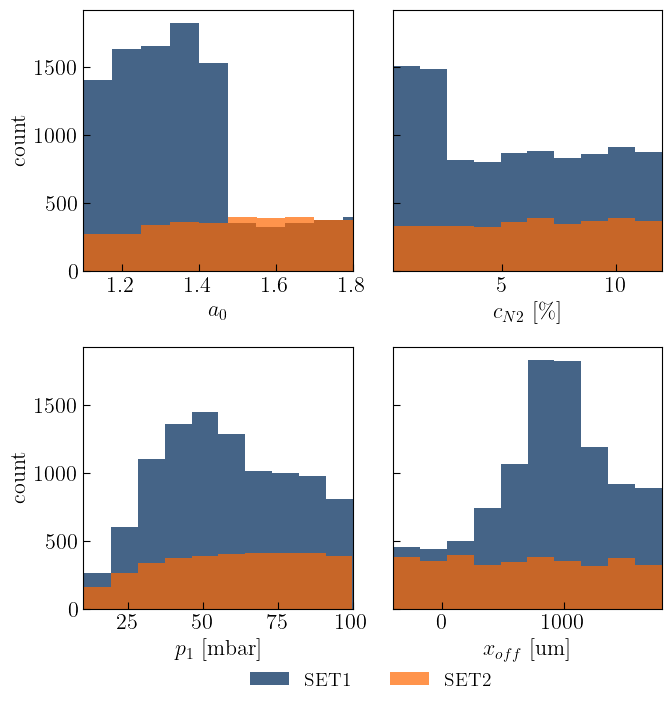}   
    \caption{Distribution of the input parameters for the $9846$ training data  simulations (blue) and the $3536$ test data simulations (orange).}
    \label{fig:inputDistribution}
\end{figure}

We chose 4 output parameters to characterise the electron beam, namely: $ y^{(j)} = (E_{med}, \delta E_{mad}, Q, \epsilon_y)$. $E_{med}$ is the median energy, $\delta E_{mad}=\sigma_{mad}/E_{med}$ with $\sigma_{mad}$ the median absolute deviation, $Q$ the charge and $\epsilon_y$ the transverse normalised emittance. The laser driver was linearly polarised along the $y$-axis. The output parameters $y^{(j)}$ used as validation data in the model were evaluated only at the last time step of the simulation. They represent the features of the beam right after the plasma outramp. For additional details on the output beam parameters evaluation see Appendix A.
The first simulations data set \textsc{set1} consists of five massive random scans, each with $2401$ configurations. Each random scan explored a part of the input parameter space using either a continuous uniform distribution or a skew-normal distribution. These random scans resulted in some intervals of the input space being over-represented in the data set. The histograms in Fig.~\ref{fig:inputDistribution} present the configurations distribution of the inputs $x^{(i)}$. 

The density of points within the random scan data set \textsc{set1} was high enough to explore the input parameter space finely. The resolution is largely above the one that can be obtained experimentally. For example, with $x_{off}$, we reach a numerical resolution as low as $1\,\mu$m where it is barely  $50\,\mu$m in standard experimental conditions.

\begin{table}[!h]
 \centering
\begin{tabular}{|l|c|c|c|c|}
\hline
data set & $E_{med}$ [MeV] & $\delta E_{mad}$ [\%] & $Q$ [pC]& $\epsilon_y$ [mm.mrad]\\
\hline
\textsc{set1} & $[41,355]$ & $[0.01,58]$ & $[3,791]$ & $[0.6,78]$ \\
\hline
\textsc{set2} & $[44,368]$ & $[0.01,56]$ & $[3,837]$ & $[0.6,76]$ \\
\hline

\end{tabular}
\caption{Intervals for the four output electron beam parameters across the simulation data sets.}
\label{tab:outputParameters}
\end{table}


A second simulations data set \textsc{set2} was produced using an injection prediction model (see \ref{subsec:injection}) for filtering input parameters $x^{(i)}$ resulting in $3536$ simulations.  For the \textsc{set2} data, the input parameters $x^{(i)}$  were randomly drawn from the intervals presented in Tab. \ref{tab:inputParameters} using a continuous uniform distribution (Fig.~\ref{fig:inputDistribution}). 

For the \textsc{set1}, out of the $12004$ simulations, $9846$ resulted in injected beams. The output $y^{(j)}$ ranges are presented in Tab.\ref{tab:outputParameters} for both data set \textsc{set1} and \textsc{set2}. 
The two data sets were used both as training data or test data.

\subsection{Injection prediction model}\label{subsec:injection}

We constructed an injection model trained on the initial data set \textsc{SET1}. This model predicts whether injection will happen for the input $x^{(i)}$. It uses a simple random forest algorithm to make its prediction. The accuracy of the model is $98$\%. Accuracy denotes the number of correct predictions over the total number of predictions. This injection prediction model can save computational time before launching new simulations or be used as a constraint in the Bayesian optimisation search to filter the input parameters.

\begin{figure}[!htp]
    \centering
    \includegraphics[width=0.5\columnwidth]{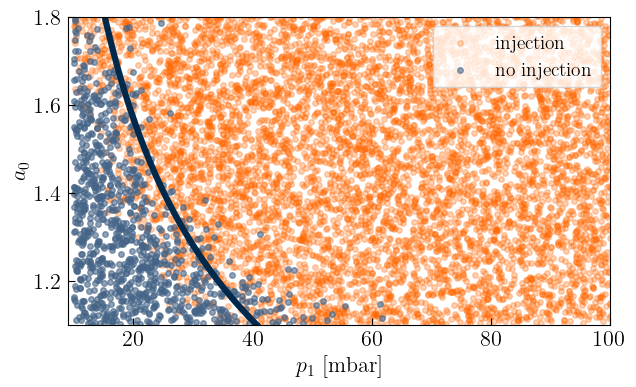}
    \caption{Injection model tested on randomly generated points. The dark-blue curve is $2.5 \cdot p_c$, with $p_c$ the critical pressure necessary for self-focusing.}
    \label{injection_model}
\end{figure}

Fig.~\ref{injection_model} shows, as a function of the input parameter ($a_0$,$p_1$): in blue, the points corresponding to no injection, in orange injection. The definition of injection is an accelerated beam ($\gamma_z>10$) with charge $Q>3\,$pC. The black curve corresponds to $2.5 p_c$, with $p_c$ the critical pressure  for self-focusing as defined in \cite{drobniak2023fast,lu2006generating}.
\begin{equation}
    p_c \simeq 16 \cdot \frac{\epsilon_0 k_B T m_e c^2}{e^2 w_0^2 } \cdot \frac{1}{a_0^{2}} \cdot \frac{1}{(1+4 c_{N_2}) }
\end{equation}
with $w_0$ laser waist, and normalised vector  potential $a_0$. 
It can be seen that the injection condition determined by the model is proportional to the theoretical limit set by the self-focusing threshold. 
However, electron injection by ionisation from N$^{5+}$ and N$^{6+}$ limit is greater than the critical pressure for self-focusing, given the limited intensity assumed for the laser driver pulse in the study, $a_0 \in [1.1,1.85]$.

\section{Model construction}

The current section intends to present the training process of the models and the importance of the input parameter distribution in the training data set.
\subsection{LPI model}

We study and compare various ML methods and eventually determine which one would be the most appropriate to model and optimise the LPI. The methods considered are:
\begin{itemize}
\item  Neural network (NN) like multilayer perceptron (MLP) – which is a well-generalised robust method for learning nonlinear data \cite{jalas2021bayesian}.
\item Trees like Extreme gradient boosting (XGB)– a classical method for learning by splitting data into different branches. These methods are fast but tend to overfit \cite{dietterich1995overfitting}.
\item Gaussian Processes (GP) - a statistical method allowing the prediction of the expected value and its variance. It has no hyper-parameters to tune after the Kernel and length are defined. It is at the core of Bayesian Optimisation, which has been successfully used in multiple accelerator physics applications and studies \cite{jalas2021bayesian}.
\end{itemize}


\subsubsection{multilayer perceptron (MLP)}

The first method used to construct a surrogate model of the LPI consisted of four different MLP's. These MLP's were implemented using \textit{Tensorflow} and \textit{Keras} python library \cite{tensorflow2015-whitepaper}. Each MLP predicts one output parameter. For $E_{med}$, $Q$ and $\epsilon_y$. The MLPs have the following architecture: $5$ layers in total, $1$ input layer with $4$ neurons, $1$ output layer with $1$ neurons, and $3$ intermediate layers with $100$ neurons. Each layer has a $20\%$ dropout rate. We used the function PRELU \cite{he2015delving} as an activation function for the $3$ intermediate layers and a sigmoid function for the last layer. Altogether, this model contains $21101$ trainable parameters. This model was trained on $200$ epochs with a batch size of $50$. The loss function used was the mean squared error (MSE). To avoid over-fitting, we also used a \textit{K-fold} cross-validation method \cite{refaeilzadeh2009cross}. For predicting  $\delta E_{mad}$ the architecture is modified to improve accuracy, since this parameter is highly correlated with energy and charge. The input layer contains $6$ neurons instead of $4$. These additional input neurons are the prediction of $E_{med}$ and $Q$ from the already trained MLP.

\subsubsection{Extreme gradient boosting (XGB)}

We implemented this method by using the \textit{xgboost} library\cite{chen2016xgboost}. The maximum tree depth was set to $10$, and the loss function was also MSE in this case. We used \textit{K-fold} cross-validation.

\subsubsection{Gaussian Process (GP)}

We implemented this method by using the \textit{Scikit-learn} library \cite{pedregosa2011scikit}. The kernel used was \textit{Matérn}, which is a generalisation of the Gaussian radial basis function and allows the capture of physical processes due to double differentiability by the choice of a smoothness parameter $\nu = 2.5$.

\smallbreak
The training process on a high-performance laptop is relatively quick, and it takes only a few seconds for the XGB model and a few minutes for the GP and MLP models. Additionally, the computation time for LPI configurations is significantly shorter than that of low-fidelity simulations on HPC CPU nodes. The MLP, XGB, and GP models are approximately $10^{7}$ , $10^{8}$, $10^{7}$ times faster than simulations, respectively. The time taken for training and inference are presented in Appendix B. It is important to note that for all models (MLP, XGB, GP), it is mandatory to rescale the outputs and the inputs from $0$ to $1$ to get the most accurate results. The output parameters, $y^{(i)}$, are scaled so that the calculation of the loss function is well-weighted, corresponding to the same magnitude in all of the outputs.

\smallbreak

\subsubsection{Importance of the input parameters distribution}

All three models MLP, XGB, and GP were tested on the \textsc{set2} data, consisting of 3536 test points, separate from the 9846 samples of \textsc{set1} used for training. 
We observe that the coefficient of determination $R^{2}$ between the SM predictions and the outputs of \textsc{set2} is above 0.85. However, this score significantly decreases in regions where the density of training points is lower than 1. The density is defined as the number of points inside a hypercube of a side $0.1$ in the normalised hyperspace. 

We propose a reliability criteria for the SM models based on the relation between $R^{2}$ and $MSE$ :


\begin{equation}
\begin{aligned}
    R^{2}=1-\frac{\sum_{i=1}^{4}\sum_{j=1}^{N}(y_{ij}-f_{ij})^{2}}{\sum_{i=1}^{4}\sum_{j=1}^{N}(y_{ij}-\bar{y}_{i})^{2}}=\\ 1-\frac{4\sum_{j=1}^{N}MSE_j}{\sum_{i=1}^{4}\sum_{j=1}^{N}(y_{ij}-\bar{y}_{i})^{2}}
\end{aligned}
\end{equation}

With $y$: output value of the test data, $f$: predicted value by the surrogate, $\bar{y}$: simulation mean value for a batch of size $N$. The index $i$ represents the 4 output parameters, and $j$ represents the test points. $MSE_j$ is the mean squared error for a specific test point $MSE_{j}=\sum_{i=1}^{4}(y_{ij}-f_{ij})^{2}/4$.






\begin{figure}[ht]
\centering
\includegraphics[width=0.5\columnwidth]{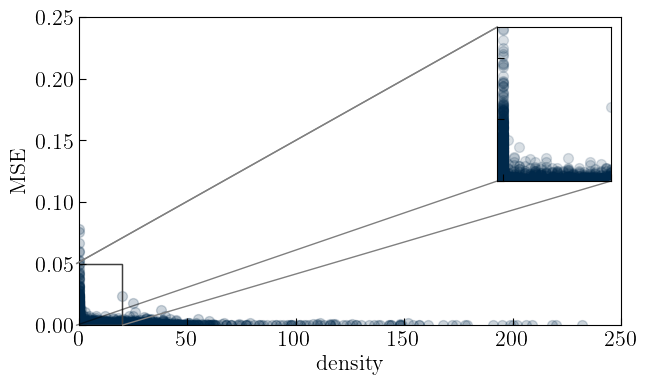}
\caption{Scatter plot showing the relationship between the density of input data configuration points and MSE for every test point for the MLP model. The blue points represent the different configurations. } 
\label{fig:densvsloss}
\end{figure}

\begin{figure}[!ht]
    \centering
      \includegraphics[width=0.5\columnwidth]{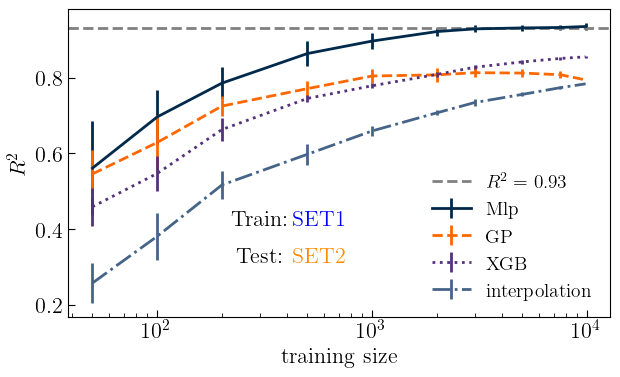}
      \caption{Coefficient of determination $R^2$ as a function of the training size in log scale, for all the SM trained on \textsc{SET1}  and tested on \textsc{SET2}. $R^2$ was taken as the average over 10 training sessions, with the vertical bars representing the standard deviation.}
      \label{fig:r2vstrainset1}
  \end{figure}

The relationship between $MSE$ and $R^{2}$ can be used to define reliability criteria for the test points in a given region of the input parameter space. 
We observe that $R^2 \geq 0.9$ for a given batch corresponds to $MSE<4.10^{-3}$ and that the probability of getting a small $MSE$ increases with the local density of training points as shown in Fig.~\ref{fig:densvsloss}. Thus, to be confident that $R^2 \geq 0.9$ in every region of the 4-dimensional input space we need to have a high enough density, which was unfortunately not the case for at least $30 \%$  of the parameter space when using the simulation data of \textsc{SET1} \cite{drobniak2023fast} as training. This is why the following models were trained with data from \textsc{SET2} uniformly distributed points and tested on the data of \textsc{SET1}. As shown in   Fig.~\ref{fig:r2vstrain}, \ref{fig:r2vstrainset1} a better homogeneity largely compensates for reducing the number of training points for our LPI SM.





\section{Results}
\subsection{Performances of surrogate models}

The SM trained on the data \textsc{set2} showed good correlation with  MLP, XGB  and GP model having an $R^{2}$ score of $0.97$, $0.90$ and $0.96$, respectively, across all output parameters. However, as shown in Table \ref{tab_correlation}, the median energy and charge are consistently better predicted by the models compared to  $\delta E_{mad}$ and $\epsilon_{y}$. To compare the results of the SM with a more standard method, we added nearest-neighbour interpolation.

\begin{table}[!h]
 \centering
    \begin{tabular}{|l|c|c|c|c|}
    \hline
     &$E_{med}$ & $\delta E_{mad}$ & $Q$ & $\epsilon_{y}$\\
    \hline
    MLP &            0.99 & 0.96& 0.99 & 0.95 \\
    \hline
    XGB &           0.97 & 0.88& 0.96 & 0.78 \\
    \hline
    GP &            0.99 & 0.95& 0.99 & 0.90 \\
    \hline
    interpolation & 0.90 & 0.83 & 0.91 & 0.72 \\
    \hline
    \end{tabular}
    \caption{$R^2$ correlation score for the different surrogate models trained on SET2 and evaluated on SET1.}
    \label{tab_correlation}
    \end{table}

\begin{figure}[!ht]
    \centering
      \includegraphics[width=0.5\columnwidth]{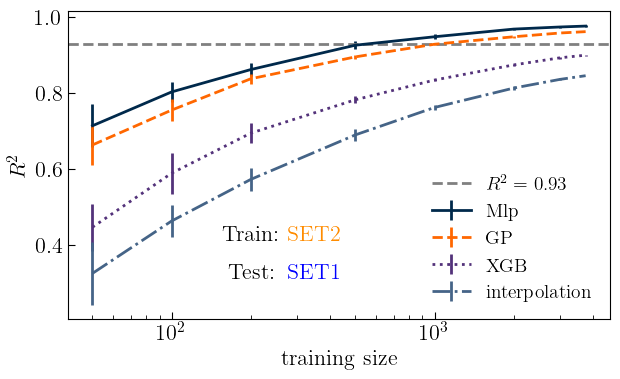}
      \caption{coefficient of determination $R^2$ as a function of the training size in log scale, for all the SM trained on \textsc{SET2}  and tested on \textsc{SET1}. $R^2$ was taken as the average over 10 training sessions, with the vertical bars representing the standard deviation}
      \label{fig:r2vstrain}
  \end{figure}

Figure \ref{fig:r2vstrain} illustrates that the MLP and GP model arrive at $R^2=0.93$  with a training size of approximately $500\,$samples. All SM outperform a simple interpolation model, which only reaches a $R^2$ score of $0.85$ when the training size exceeds 3000 samples compare to less than 300 samples for MLP and GP SM. 

To evaluate the performance of the SM across the entire output interval, we computed the mean absolute error (MAE) for each small slice of these intervals. 

\begin{figure}[!ht]%
	\centering
	\includegraphics[width=0.5\columnwidth]{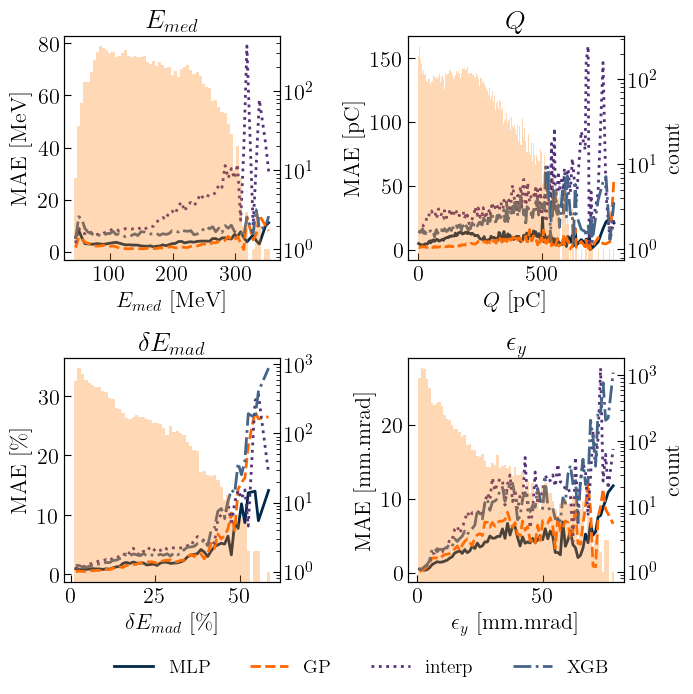}%
	\caption{MAE for all the SM and all of the output with histograms representing  the distribution of the output parameters of training data set \textsc{set2} in log scale}
	\label{fig:MAE}
\end{figure}

\begin{figure}[!ht]%
	\centering
	\includegraphics[width=0.5\columnwidth]{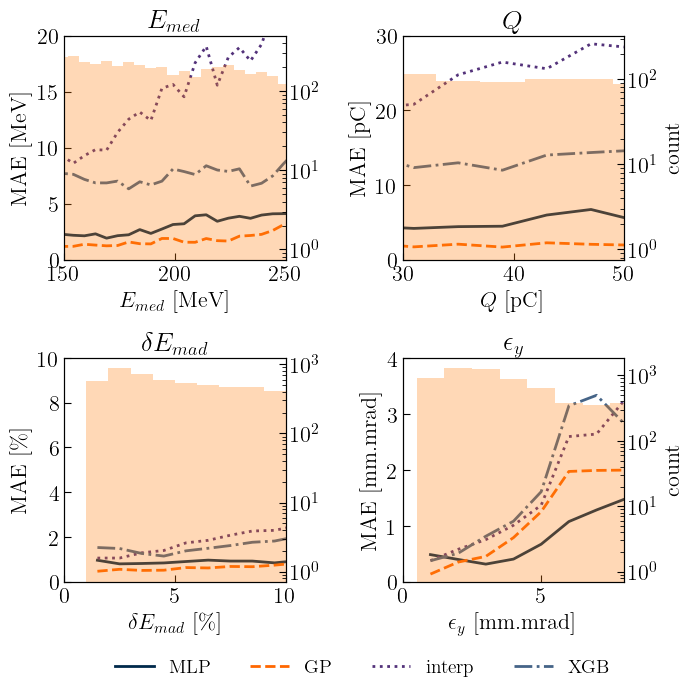}%
	\caption{MAE for all the SM and all of the output with histograms representing the distribution of the output parameters of training data set \textsc{set2} in log scale zoomed on the region of interest for outputs.}
	\label{fig:MRE}
\end{figure}

\noindent Fig.~\ref{fig:MAE} shows, that the MLP model is the best followed by the GP. 
Although the MLP model is the best overall Fig.~\ref{fig:MRE} shows that the GP model is the best in the ranges of interest ( $150-250\,$ MeV, with $30-50\,$pC of charge, an energy spread lower than $5\%$ and an emittance of less than $2\,$mm.mrad). 
\noindent Additionally, the MAE tends to increase for the highest output values since these values are underrepresented in the training data set as shown by the histograms depicting the distribution of the output parameters of \textsc{set2} in Fig.~\ref{fig:MAE}.

\begin{figure*}[!ht]
    \centering
    \includegraphics[height=\textwidth]{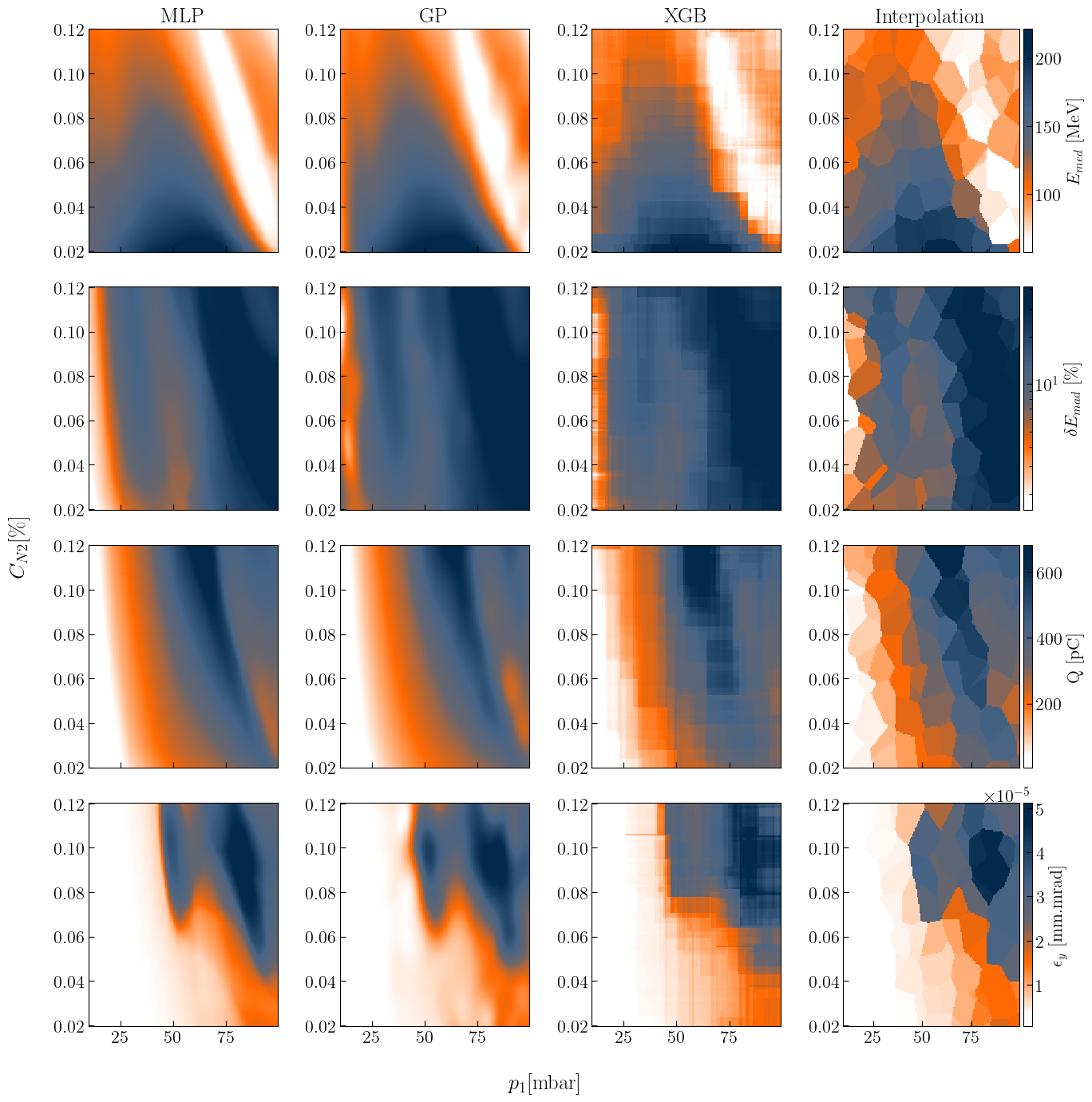}
    \caption{Surrogate LPI models prediction for all of the output in a 2D subspace of $p_1$ and $c_{N2}$ for $a_{0}=1.43$ and $x_{off}=-265\,\mu m$. Other snapshots of 2D subspace can be generated using the python notebook available on the online repository\cite{LPISurrogategitlab2024}}
    \label{fig:allModels}
\end{figure*}

\noindent In Fig.~\ref{fig:allModels}, the prediction of each SM is represented in a 2D subspace of $c_{N_2}$ and $p_{1}$. The projections are made for the input laser parameters fixed to $a_0=1.43$ and $x_{off}=-265\,\mu$m. One should notice that a complete set of projections can be generated in a few seconds on a laptop for a complete scan of $a_0$ and $x_{off}$ or other target parameters for more complex studies.   

\subsection{Optimisation with the surrogate models}

The GP model was employed in the subsequent analysis due to its superior performance compared to the XGB and interpolation methods. Although it is less accurate than the MLP across the entire output space, the GP model demonstrates higher efficiency within the specific range of interest $150–250$ MeV peak energy, $30–50$ pC charge, energy spread below $5\%$, and emittance under 2 mm·mrad—as illustrated in Fig.~\ref{fig:MRE}.


\subsubsection{Optimum LPI working point stability}

Using the GP model, we looked for optimal working points. Several methods can determine these configurations of target and laser for the optimal electron beam parameters. The simplest approach is to generate many data points using a continuous uniform distribution of 4D input parameters. The input range is kept within the boundaries of Tab. \ref{tab:inputParameters}.

\begin{figure}[!htp]
    \centering
    \includegraphics[width=0.5\columnwidth]{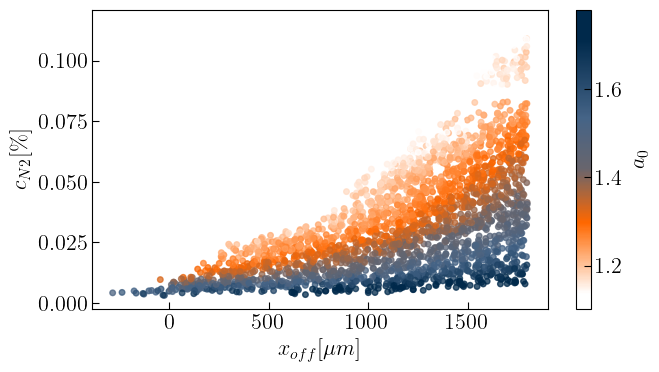}
    \caption{2D subspace of $x_{off}$ and $c_{N2}$ of the input parameters for the configurations selected by filter $\tilde{F_1}$ with $a_0$ as a color scale.}
    \label{fig:2dspacesXoffCn2}
\end{figure}

From this data set, our model can then be used to select beams with the desired characteristics. Selection is performed using the following filter: $\tilde{F_1}= \{E_{med} \in [205,215]\,$MeV, $\delta E_{mad} < 3.5\,$\%, $Q \in [25,35]\,$pC$,\epsilon_y<2 $mm.mrad$\}$. We generated $5\,$million random configurations using a uniform distribution and then used the injection model to keep only the configurations that predict injection.
From these configurations, $2347$ were selected by filter $\tilde{F_1}$. We can see in Fig. \ref{fig:2dspacesXoffCn2}, for each value of $x_{off}$, $c_{N2}$ logically decreases with $a_0$ since the target charge is fixed in the filter. If the laser energy is lower, the charge can be maintained at a certain level by increasing the doping rate, as explained in \cite{drobniak2023fast}. Not only the number of injected electrons can be increased, but also increasing the self-focusing helps to reach the threshold value for $a_0$.

\begin{figure}[!htp]
    \centering
    \includegraphics[width=0.5\columnwidth]{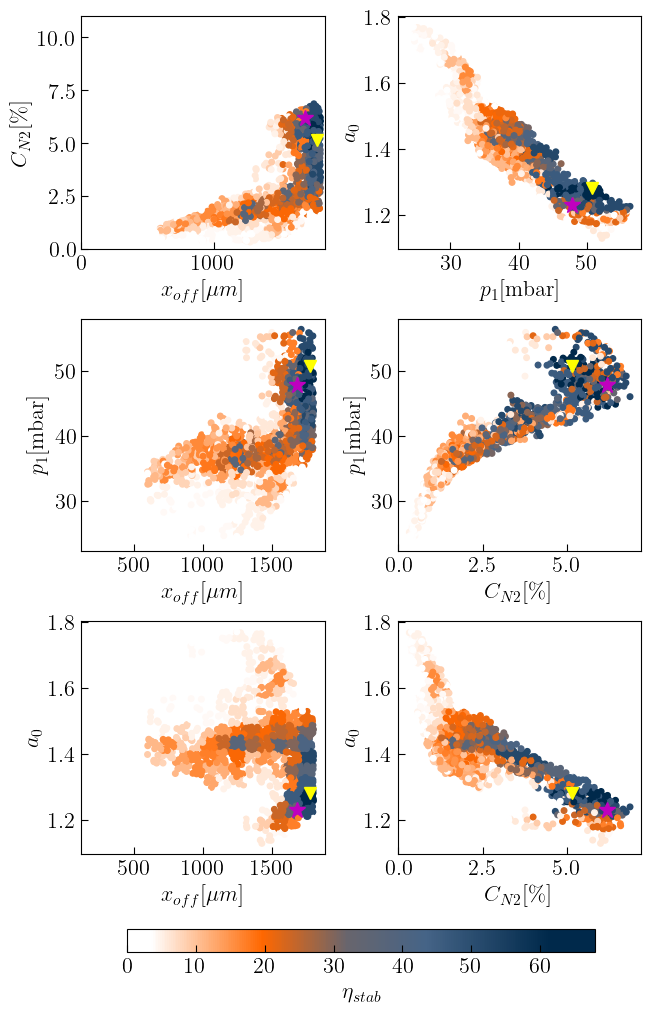}
    \caption{Stability map: Projection in the 2D sub-spaces of the configurations selected by filter $F_1$ with $\eta_{stab}$ as a color scale. These maps allow us to see the location of the most stable regions. The $\star$ symbol is the configuration 7516 from \textsc{SET1} and the $\triangledown$ symbol is most stable point in the filter $F_1$}
    \label{fig:stability_map}
\end{figure}

One interesting aspect to examine with this method is the stability across a target working point. Since the 5 million points were generated using a uniform distribution, each region of the 4D input space contains roughly the same number of points. Thus, we consider that the region with the highest density of remaining points after applying $\tilde{F_1}$ in the input space is the most stable. In Fig.~\ref{fig:stability_map}, we present stability maps as projections in 2D sub-spaces, showing the density of points $\eta_{stab}$ in the input space of filter $\tilde{F_1}$, the density is represented with the colours scale. These stability maps can guide the search for ideal electron beams. From this analysis, we identified that the most stable region is centred around the following point: $a_0=1.31$, $c_{N_{2}}=6.1\,$\%, $x_{off}=1.676\,$mm, $p_1=39\,$mbar.

\subsubsection{Bayesian optimisation}

Bayesian optimisation can also be used with SM. We can employ either single or multi-objective Bayesian optimisation (MOBO). For single-objective optimisation, we used the following function to be optimised: $\tilde{f}_3=\sqrt{Q}E_{med}/\delta E_{mad}$ \cite{jalas2021bayesian}. The optimisation consisted in one hundred steps with $20$ random evaluations. Out of the $120\,$points from the Bayesian optimisation, $39$ met or exceeded $95\,$\% of the maximum of the objective function. These configurations have an average charge of $149 \pm 6 \,$pC, an energy of $236 \pm 3 \,$MeV and energy spread of $2.8 \pm 0.05\,\%$. 

\begin{figure}[!ht]
    \centering
    \includegraphics[width=0.5\columnwidth]{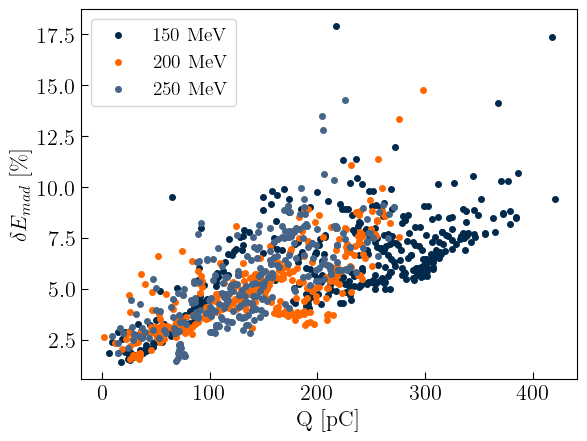}
    \caption{Scatter plot showing the result of MOBO search for 3 different energies $150,200$ and $250~$MeV within a $\pm 10$ MeV}
    \label{Paretofront}
\end{figure}

The MOBO aims at optimising simultaneously the elements of the following vector $\tilde{G}=(\delta E_{mad}(x),|E_{med}(x)-E_{0}|,Q(x))$ where $x$ is the 4D input vector. Our goal here is to maximise charge and minimise energy spread for a given median energy. We tried three MOBO searches with the vector $\tilde{G}$ for three different central median energies $150$, $200$ and $250$ MeV within a $\pm 10$ MeV. Each MOBO consisted in $80$ steps with $10$ random evaluation and $10$ evaluation for each step. This  MOBO search resulted in Pareto fronts \cite{irshad2023pareto}, illustrated in Fig.~\ref{Paretofront}.



    



Here, we can see a clear trade-off between charge and energy spread. These solutions correspond to \textit{Pareto}-optima  \cite{jalas2023tuning} as we cannot improve one of the objectives without deteriorating the other.



These three approaches permit the finding of target working points for the LPI. However, we want to emphasise that the first one with filter $\tilde{F_1}$ is the most mature one because this allows us to find the beams of interest and the most stable LPI configurations. This approach, however, requires generating a large amount of data, which is not a problem with the MLP model contrary to PIC simulations.

\subsection{Comparison with random scan optimisation}

The previous method can be used to seek configurations that outperform the best ones identified in \cite{drobniak2023fast}. In this work, two criteria were used for beam selection: $f_3=E_{med} \cdot Q /\sigma_{mad}$ and filter $F= \{E_{med} > 150\,$MeV, $\delta E_{mad} < 5\,$\%, $Q >30\,$pC$,\epsilon_y<2 \mu m\}$. The best beams were chosen based on the following: The configuration that maximised $f_3$  (configuration \textsc{3702}) from \textsc{SET1}, the configuration within filter $F$ that had the smallest $\delta E_{mad}$ (configuration \textsc{7516}) from \textsc{SET1}. The characteristics of configurations \textsc{3702} and \textsc{7516} are presented in Tab .\ref{tab_bestconfig}. To compare our method with the results from \cite{drobniak2023fast}, within the 5 million random configurations, we identified 3779 configurations with higher $f_3$ values than configuration \textsc{3702} and 408 configurations that satisfied the conditions of filter $F$ with a smaller $\delta E_{mad}$  than configuration \textsc{7516}. The results are presented in Tab .\ref{tab_bestconfig}.

\begin{table}[!h]
 \centering
    \begin{tabular}{|l|c|c|c|c|}
    \hline
    &  $f_3^{(opt)}$ \cite{drobniak2023fast} & $F_2^{(opt)}$ \cite{drobniak2023fast} & SM $f_3^{(opt)}$ & SM  $F_2^{(opt)}$\\
    \hline
    $x_off$ &  558 & 1680& -372 & 1798 \\
    \hline
    $a_0$ & 1.43& 1.23& 1.24 & 1.33 \\
    \hline
    $C_{N2}$ & 1.88 & 6.17& 9 & 7.70 \\
    \hline
    $p_1$ & 58.6& 47.8& 88 & 40.7 \\
    \hline
     $E_{med}$ & 215& 212 & 103 & 185 \\
    \hline
    $\delta E_{mad}$ & 3.53& 1.55& 3.09 & 0.9 \\
    \hline
    $Q$ &  198 & 30& 311 & 45 \\
    \hline
    $\epsilon_y$ & 5.03& 1.74 & 38 &1.5 \\
    \hline
    \end{tabular}
    \caption{Input and output parameters of the best configurations found by the random scan and filter}
    \label{tab_bestconfig}
\end{table}

\begin{figure}[!h]
    \centering
    \includegraphics[width=0.5\columnwidth]{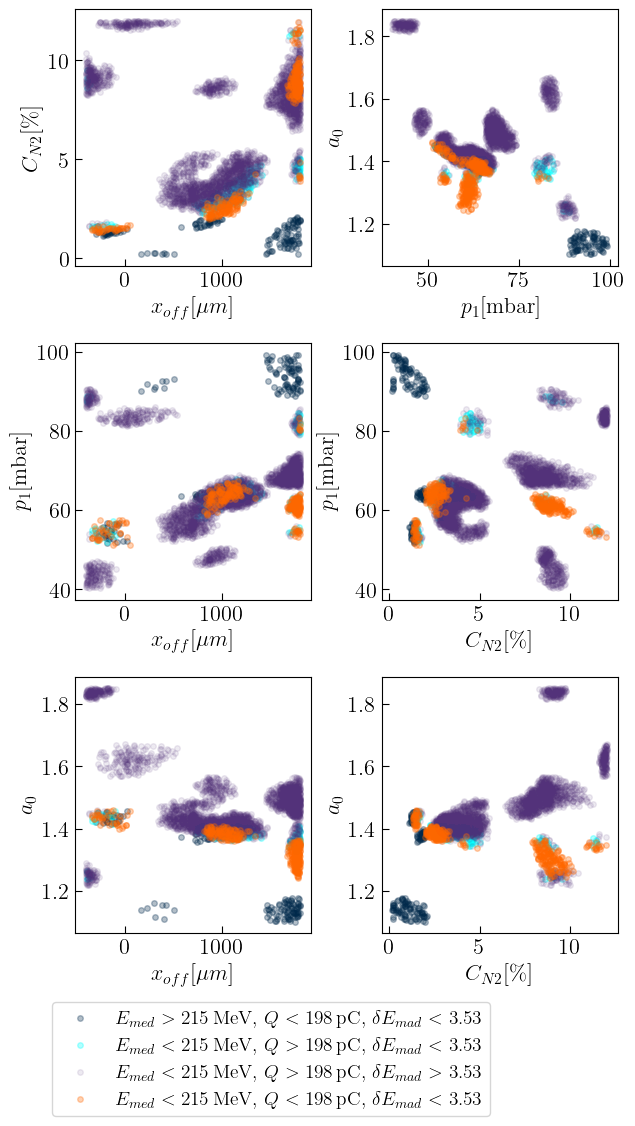}
    \caption{2D subspaces of the configurations selected by the function $f_3$ higher than 3702 with regions of interest highlighted by different colours.}
    \label{fig:2dspacesofinterests}
\end{figure}

For configuration SM $f_3^{(opt)}$, we have an upstream focus coupled with a high pressure in target chamber 1 and high dopant concentration, leading to strong self-focusing; all of which leads to a high charge beam even with relatively low intensity. However, the high amount of charge leads to a high emittance value. For SM $F^{(opt)}$, we have a downstream focus with moderate pressure and intensity, which leads to a low injected charge where very low energy spread is possible. We can thus see that the SM can find working points that outperform a simple random scan. In addition, it is interesting to look at regions of interest, for example, we show in Fig. \ref{fig:2dspacesofinterests}. the 2D sub-spaces of all the points that have $f_3$ values superior to configuration \textsc{3702} and identify regions of interest with different colours. 
At first glance, the different beams are clustered in different areas. A first group of interest is $E_{med}>215,\delta E_{mad}<3.53 \%$.  Most of this set is located at high $x_{off}$ and high $p_1$ values, with $x_{off}$ above 1500 $\mu m$. The combination of high pressure and downstream focus leads to a more important $a_0$ in the accelerating region, leading to a higher wakefield amplitude and high energy beams. This set also displays low $c_{N2}$ and low $a_0$, which limits the amount of injected charge to reasonable values ($58$ to $94\,$pC) despite the high pressure. The reasonable electron bunch charge also allows for maintaining small energy spread by limiting space charge and unwarranted beam loading. The most prominent group is the high charge case $Q>198\,pC$. This set is distributed all over the hyperspace but we find that it generally follows this guideline the higher $a_0,C_{N2},p_1$ the higher the charge will be. Increasing any of these three parameters means increasing the number of inner-shell ionised electron, all other things being equal. The lower $x_{off}$, the higher the charge will be following the trends of a lower $x_{off}$  balances a higher $a_0$ in the injection region and thus a higher inner-shell ionisation rate. 

\section{Conclusion and perspective}

In conclusion, our in-depth numerical study focused on applying machine learning in the context of a laser plasma injector optimisation design. It demonstrates that a surrogate model approach is relevant for beam optimisation and stability, increasing efficiency and requiring a lower computational cost. We successfully constructed models that exhibit high performance in predicting electron beam parameters.

We emphasised the importance of data distribution in achieving accurate results with SM. Our analysis has shown that $R^2$ scores converge rapidly towards $1$ if trained on sufficiently uniform data sets. This capability enabled us to use the surrogate models effectively to identify optimal working points for designing LPI  electron sources.

Furthermore, SM provides a comprehensive view of potential LPI beam parameters, facilitating the identification of stable operational regions and can drive the development of plasma targets for higher repetition rate laser-plasma accelerators with limited laser intensity. These models are straightforward to implement and can be continuously refined by incorporating new simulation data.

However, we identified certain limitations of the surrogate models. They tend to underperform in regions where data points are sparse and exhibit poorer performance at the lower end of the output range (low charge, low emittance and low energy spread). A potential improvement could be to add new simulation data in those regions and train our model on a larger interval of the output space than the test region.

Our study highlighted the ability of MLP and GP to generalise well and achieve the highest predictive performance among the ML methods considered. The LPI surrogate model can be used as an electron source for start-to-end simulation studies, opening the way to model a full accelerator beamline with variation in LPI electron source input parameters.

These promising results show that these methods could eventually be used with experimental data of the LPI or a hybrid version between experimental and simulation data since the time necessary to gather a large amount of experimental data is much shorter than PIC simulations. Such models could be implemented using a multi-fidelity approach as in \cite{irshad2023multi}. More weight would be added to experimental data in comparison to simulations.


The next step would be to apply the SM within a control framework to introduce an efficient feedback loop for LPI beam stabilisation and control. This could be achieved either by using a reverse SM \cite{bethke2021,miethlinger2023} to map the output space back to the input space, although this task is numerically challenging since the four output parameters are not independent, and the existence and uniqueness of the solution are not guaranteed, or by embedding the forward SM into a model predictive control scheme \cite{schwenzer2021review} or a model-based reinforcement learning framework  \cite{moerland2023model}, both of which can directly exploit the predictive power of the SM to optimise control actions.

\section{Data Availability Statement}
The features, data, and models supporting this study's findings are available online \cite{LPISurrogategitlab2024}. Raw PIC simulation data are available from the corresponding author upon reasonable request. 


\section{Acknowledgments}
This work has benefited from European funding EUPRAXIA-PP HORIZON-INFRA-2021-DEV-02 EUR project 101079773.
This work was granted access to the HPC resources of TGCC Irene Joliot Curie under the allocations 2021 - A0110510062 and 2022 - A0130510062 made by GENCI for the project Virtual Laplace.



\subsection*{Appendix A: Output electron beam parameters evaluation}
 
\begin{figure}[!ht]
    \centering
    \includegraphics[width=0.5\columnwidth]{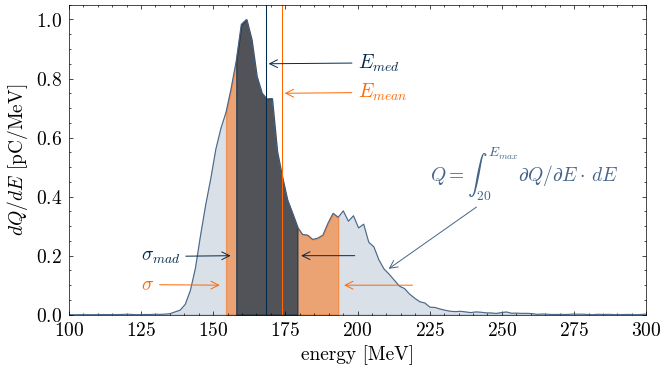}   
    \caption{electron beam statistical features processed from the energy distribution}
    \label{fig:test}
\end{figure}

 The electron beam features are retrieved using a postprocessing script based on \textsc{APtools} python code \cite{aptools2024} for the last time step corresponding to the end of the electron plasma density longitudinal profile. The figure Fig.~\ref{fig:test} shows an example of median energy $E_{med}$, median absolute deviation $E_{mad}$ and charge corresponding to the integration of the electron beam energy distribution from the minimum energy tracked in the PIC code to the maximum energy.  

Only the trackParticles SMILEI openPMD\cite{openpmd} output is used to post-process electron beam features. The method used in the present paper can be extended to any PIC code data using openPMD format. 

\subsection*{Appendix B: Training surrogate model construction resources requirement}

The training of the LPI surrogate models, as the number of input and output is limited, was done using a standard \textsc{I7} Intel cpu. 

The following table tab summarises the SM training time and computing time to generate $10^{5}$ LPI configurations. \ref{tab:smresources}.
\begin{table}[!h]
 \centering
    \begin{tabular}{|l|c|c|}
    \hline
    time [s] & training & computing $10^{5}$ configurations \\
    \hline
    MLP &            850 &  10   \\
    \hline
    GP &            52 &  14  \\
    \hline
    XGB &            13 & 0.34  \\
    \hline
    \end{tabular}
    \caption{Training time and computing for each type of method used to build the LPI model using \textsc{I7} Intel cpu.}
    \label{tab:smresources}
    \end{table}


\input{iopjournal-template.bbl}


\end{document}

%% file: iopjournal-template.bbl
\providecommand{\noopsort}[1]{}\providecommand{\singleletter}[1]{#1}%